      [1999/12/01 v1.4c Il Nuovo Cimento]
\documentclass{cimento}
\usepackage{graphicx}

\title{Measuring Star Formation in Local and Distant Galaxies}
\author{D.~Calzetti\from{ins:x}
}
\instlist{\inst{ins:x} Department of Astronomy, University of Massachusetts, Amherst, MA, USA
                          }
\PACSes{\PACSit{95.85.-e}{Astronomical Observations; Multiwavelength}
\PACSit{98.52.Nr}{Spiral Galaxies}
\PACSit{98.52.Sw}{Irregular and Morphologically Peculiar Galaxies}
\PACSit{98.54.Ep}{Starburst Galaxies}}
\begin{document}

\maketitle

\begin{abstract}
I review measurements of star formation in nearby galaxies in the UV--to--FIR wavelength 
range, and discuss their impact on SFR determinations in intermediate and high redshift 
galaxy populations. Existing and upcoming facilities will enable precise cross-calibrations 
among the various indicators, thus bringing them onto a common scale. 
\end{abstract}

\section{Introduction}
Determinations of star formation rates (SFRs) in galaxies utilize  
indicators at a variety of wavelengths, from the X--ray to the radio. Many 
indicators have been defined in response to specific needs. For instance, 
when new populations of galaxies are discovered using a new  
wavelength window or improved observing techniques/instruments in a 
certain waveband, there is a push to investigate whether that waveband 
can be used to derive SFRs as well, and/or to define the uncertainties and 
limitations of doing so. 

The advent of new facilities (e.g., Herschel, LMT, ALMA, JWST, etc., and the many 
ground--based telescopes under construction or design) together 
with existing ones (HST, Spitzer, Chandra, and the vast array of existing ground--based 
facilities) will cover extensively the electromagnetic spectrum at unprecedented sensitivities. 
This will offer the opportunity to cross-calibrate many of the SFR indicators across a 
range of redshifts, and, therefore, on many galaxy populations at various 
stages of their evolution. 

In this brief review, I discuss the current status and the known limitations of 
SFR indicators in a few wavelength regimes: ultraviolet (UV), optical/near--infrared, 
and mid/far--Infrared (MIR/FIR). 

\section{General Assumptions and Limitations}
By definition, most SFR indicators probe the {\em massive stars formation rate}; their 
scope is  to measure the instantaneous or recent SFR of a galaxy or system 
(as opposed to the time--averaged SFR). Thus, the measured  luminosity L($\lambda$) 
needs an assumption on the stellar Initial Mass Function (IMF) to be converted to an actual 
SFR. All indicators are insensitive to the low-end of the stellar IMF, which thus remains 
a free parameter (or an uncertainty) for all such measures. The sensitivity 
to the high--end of the IMF varies from indicator to indicator (as we will see in the next 
sections), which thus complicates comparisons among different indicators. The issue of the 
form and mass limits of the stellar IMF will not be discussed here, but needs to be kept in mind 
when deriving SFRs from any luminosity measurement.  

Another factor to account for is the impact of dust on the luminosity L($\lambda$) of a 
galaxy.  UV, optical, and near--infrared luminosities probe the stellar light that emerges from 
galaxies unabsorbed by dust; thus, for these wavelengths, the main problem is to 
correct the observed luminosities for the effects of dust attenuation. Infrared luminosities 
measure  the stellar light that has been reprocessed by dust and emerges beyond a 
few $\mu$m. In this case, the main problem is to establish whether the re-processed light comes 
from young, massive stars (associated with the current star formation) or from older stellar 
populations. 

In general, L($\lambda$) is the sum of the contributions from 
all its stellar populations (or from the dust re-processed light of all populations). 
Thus, deriving a SFR from L($\lambda$) implies quantifying the 
impact, if any, of any stellar population that is contributing to L($\lambda$), but is not 
part of the current star formation event. Specific examples include evolved (aged) 
stellar populations. Added to this is the well--known age--dust degeneracy, for which 
a young, dusty population observed at UV--optical wavelengths can mimic the colors 
of an old, dustless population. 

\section{SFR(UV)}
The ultraviolet ($\lambda\sim$912--3000~\AA) probes directly the bulk emission from the 
young, massive stars, thus 
could be considered the SFR indicator {\em par excellence}. Ease of access of the 
restframe UV emission (redshifted into the optical bands) for high redshift galaxies 
\cite{mad96,ste99,gia04,bou07} has sparkled, over the past dozen of years or so, extensive 
investigations on its use and limitations as a SFR estimate \cite{ken98,sal07}. 

However, a number of effects limit the use of SFR(UV), unless those effects are properly 
treated. The UV is heavily impacted by dust attenuation: A$_V$=1~mag implies 
A$_{1500\AA}\approx$3~mag, and the exact value depends on the details of the dust 
geometry in the galaxy. The star formation history of a galaxy also determines its 
UV luminosity: a system which has been forming stars at a constant level of 1~M$_{\odot}$~yr$^{-1}$ 
over the past 100~Myr is indistinguishable from an instantaneous 4$\times$10$^8$~M$_{\odot}$ burst of star formation which has been passively evolving for the past 50~Myr. Indeed, the UV probes 
star formation over the timescale in which stars are bright in the non--ionizing UV 
wavelength range, of--order 100~Myr. The age--dust degeneracy is potentially a problem when 
observing the UV colors of galaxies, as a 10$^6$~M$_{\odot}$, 300~Myr old burst has a UV spectral 
slope that is virtually indistinguishable from that of a constant star formation system forming stars at 
a rate of 1~M$_{\odot}$~yr$^{-1}$  and reddened by a color excess 
of E(B$-$V)=0.4 (although the latter is about 200 times brighter than the former at 1500~\AA, according to the Starburst99 models \cite{lei99}). 

Despite all the problems listed above, starburst galaxies follow a well defined relation between 
dust reddening and dust attenuation \cite{meu99,cal00}, in the sense that a measurement of 
UV slope or colors can be effectively used to derive the total amount of dust opacity affecting the 
system \cite{cal94}. Here starbursts are defined as systems with specific SFRs (i.e., SFR/area): 
$\Sigma_{SFR}>$0.3--1~M$_{\odot}$~yr$^{-1}$~kpc$^{-2}$. The relation between dust reddening 
and attenuation in starbursts has a relatively small dispersion around the mean trend (about 
a factor 2, Figure~1), and has been effectively used to recover the intrinsic UV emission of 
strongly star--forming systems at high redshift, such as Lyman Break Galaxies \cite{ste99,gia04}. 

\begin{figure}[ht]
\includegraphics[scale=0.4]{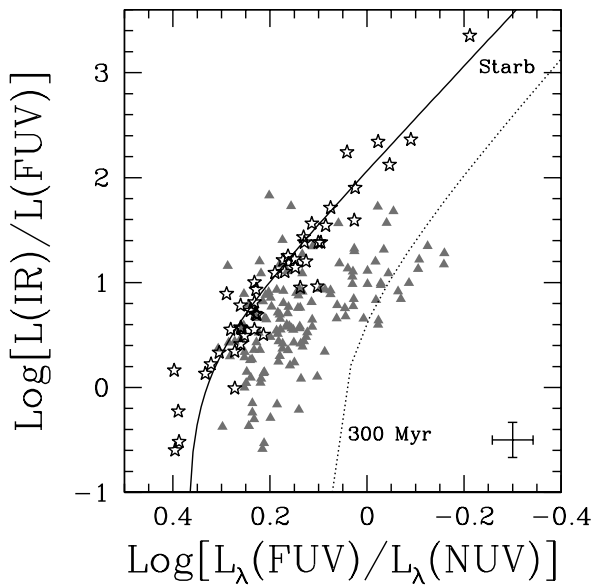}     
\caption{The FIR/UV ratio (the ratio of the far--infrared to the far--UV luminosity, a measure of dust 
attenuation) versus the UV~color (given here as the ratio between the GALEX far--UV and near--UV 
fluxes, a measure of dust reddening) for starburst galaxies and quiescently star--forming regions. 
Starburst galaxies are shown as star symbols and star--forming regions within the 
galaxy NGC5194 \cite{cal05} are shown as grey filled triangles. Redder UV colors (more dust 
reddening) correspond on average to larger FIR/UV ratios (larger dust attenuation). The continuous 
line shows the best fit to the starburst galaxies, which is also the locus of a progressively 
more attenuated (from left to right) constant star--forming population. The dotted line shows the 
same dust attenuation trend for a 300~Myr old stellar population, which represents a lower envelope 
to the NGC5194 star--forming regions.}
\end{figure}

Deviations from the well-behaved starburst attenuation--reddening relation have been observed, 
however, for other galaxy populations. Ultraluminous Infrared galaxies, for instance, show an 
excess attenuation for the measured UV reddening \cite{gol02}, likely due to the higher dust 
opacities affecting those systems. Quiescently star--forming galaxies and regions (systems with 
$\Sigma_{SFR}\ll$0.3~M$_{\odot}$~yr$^{-1}$~kpc$^{-2}$) show a ten times larger spread 
in the dust attenuation at fixed UV reddening relative to starbursts (Figure~1); the sequence for 
starburst galaxies forms the upper envelope to the distribution of the quiescently star--forming 
systems \cite{bua02,bel02,gor04,kon04,sei05,cal05}. The behavior of the quiescently 
star--forming systems is explained if the observed UV colors are not only a probe of dust 
reddening, but also of age reddening, due to contribution to the UV from aged 
(non--star--forming) stellar populations. This adds a second parameter that complicates the 
definition of SFR(UV) for these systems. 

\section{SFR(Optical) and SFR(near--infrared)}
At optical and near--infrared wavelengths, the continuum stellar emission is the result of the 
contributions from stellar populations born throughout the entire history of the galaxy. Thus, 
indicators of current/recent SFR cannot use continuum measurements. They instead rely on 
ionizing photon tracers \cite{ken98b,gal89,kew04,mou06}: the multitude of hydrogen 
recombination lines (most often H$\alpha$, H$\beta$, P$\beta$, P$\alpha$, Br$\gamma$) and 
of forbidden lines (chiefly [OII] and [OIII]). These, thus, trace the most massive, ionizing stars, and 
timescales of about 10~Myr, i.e. the most recent events of star formation in the galaxy. 

Line emission SFR indicators are more sensitive to variations of the upper end of the stellar 
IMF than SFR(UV). For comparison, a change of a factor 3 in the upper mass value of the IMF 
changes the calibration of SFR(line) by twice as much as SFR(UV). Hydrogen recombination 
lines also need to be corrected for underlying stellar absorption \cite{ros02}, and metallicity 
and ionization conditions need to be taken into account for metal forbidden lines \cite{mou06}. 

Dust extinction affects most the bluer lines: for instance neglecting extinction corrections in a 
generic sample of nearby galaxies will yield about a factor 3 underestimate in the SFR from 
H$\alpha$ \cite{ros02}. Furthermore, the underestimate will be higher  
for brighter galaxies, because of the extinction--SFR correlation \cite{cal01}. Near--infrared 
hydrogen recombination lines (P$\beta$, P$\alpha$, Br$\gamma$) would then appear to be an 
obvious choice for determining SFRs of galaxies. For example, a visual extinction A$_V$=5~mag 
(a factor 100, typical of the central regions of spiral galaxies) is reduced to A$_{P\alpha}$=0.7~mag 
(or a factor 2). However, observational limitations of ground--based telescopes, specifically the 
high atmospheric background, have so far confined the use of P$\beta$(1.28~$\mu$m) and 
Br$\gamma$(2.16~$\mu$m) to the brightest regions of galaxies (P$\alpha$(1.88~$\mu$m)  lies in 
a wavelength region where the atmospheric transmission is very low and variable, thus it is 
virtually inaccessible from the ground). 

From space the background in the H--band is about 800 times lower than from the ground, 
implying higher sensitivities, and lines like P$\alpha$ become accessible. P$\alpha$ is preferable 
over P$\beta$ and Br$\gamma$, because it is 3 and 10 times stronger, respectively.  In addition, 
unlike P$\beta$, it is unaffected by neighboring contaminating lines and is about a factor of 2 less 
sensitive to dust extinction. Current (NICMOS) and upcoming (WFC3) instruments on HST can 
access low--redshift P$\alpha$ and P$\beta$, but the small field--of--view inhibits observations of 
large or complete samples of galaxies. Future facilities like NIRSpec on JWST can access P$\alpha$ 
in galaxies up to redshift $\sim$1.5, i.e., roughly up to the peak of the cosmic SFR \cite{hop06}, and, according to models, at the peak of the dust opacity as well \cite{pei99,cal99}. This, 
in turn, calls for a term of comparison at z=0, i.e., for large samples of nearby galaxies observed in 
P$\alpha$, which can them be directly compared to the higher redshift observations.  

For an extinction--corrected hydrogen line luminosity, the calibration to SFR is given by:
\begin{equation}
SFR(M_{\odot}~yr^{-1}) = 5.3 \times 10^{-42} L(H\alpha) (erg~s^{-1}) = 4.2\times 10^{-41} L(P\alpha) (erg~s^{-1}), 
\end{equation}
for an IMF consisting of two power laws: slope $-$1.3 in the range 
0.1--0.5~M$_{\odot}$ and slope $-$2.3 in the range 0.5--120~M$_{\odot}$ \cite{kro01}. 

\section{SFR(MIR) and SFR(FIR)}

The Spitzer Space Telescope has recently enabled the investigation of the mid--infrared 
emission ($\lambda\sim$5--40~$\mu$m) as a SFR indicator, thus expanding on 
the work pioneered by ISO \cite{rou01, for04, bos04}. The interest in the MIR region stems from 
the consideration that the dust heated by hot, massive stars can have high temperatures and will thus 
emit at short infrared wavelengths. The MIR continuum is 
due to dust heated by a combination of single--photon and thermal equilibrium processes, with 
the latter becoming more and more prevalent over the former at longer wavelengths. The 
MIR bands are generally attributed to Polycyclic Aromatic
Hydrocarbons \cite{leg84,sel84}, large molecules transiently heated
by single UV and optical photons in the general radiation field of galaxies or near B stars
 \cite{peet04}, and which can be
destroyed, fragmented, or ionized by harsh UV photon fields \cite{bou88,pet05}. 

An analysis of the 24~$\mu$m (Spitzer/MIPS 24~$\mu$m band)  emission from nearby 
star--forming regions and starburst galaxies shows that this band is a good SFR tracer, 
in the absence of AGNs \cite{cal07}. A calibration can be provided, over a luminosity range 
of $>$3.5~dex:
\begin{equation}
SFR(M_{\odot}~yr^{-1}) = 1.24\times 10^{-38} [L(24~\mu m) \ (erg~s^{-1})]^{0.88} .
\end{equation}
An even better SFR tracer can be provided by combining the 24~$\mu$m luminosity (which probes  
the dust--absorbed star formation) and the {\em observed} H$\alpha$ luminosity (which probes the 
unobscured star formation):
\begin{equation}
SFR(M_{\odot}~yr^{-1}) = 5.3\times 10^{-42} [L(H\alpha)_{obs} + (0.031\pm0.006) L(24~\mu m)]. 
\end{equation}

The 8~$\mu$m emission from the same star--forming regions and starburst galaxies is, 
instead, dependent on both metallicity and star formation history, to a level that it is 
unclear whether it can be effectively used as a SFR indicator for a generic galaxy population 
(unless the basic characteristics of this population are known). 

SFR(FIR)($\lambda\sim$5--1000~$\mu$m) 
 has been calibrated since the times of the IRAS satellite, under the baseline 
assumption that, at least locally, young star--forming regions are dusty and the dust absorption 
cross--section peaks in the UV, i.e., in the same wavelength region where young, massive stars 
emission also peaks. This assumption, however, has been known to be approximate for at 
least as long \cite{hun86,per87,row89}.  The first approximation is related to the opacity of 
a galaxy: not all the luminous energy produced by recently formed stars is
re-processed by dust in the infrared; in this case, the FIR only recovers part of the SFR, and 
the fraction recovered depends, at least partially, on the amount of dust in the system. The second 
approximation is related to the heating of the dust by evolved, non--star forming population: these 
will also contribute to the FIR emission, providing an excess to SFR(FIR). If more evolved populations 
contribute mainly to the longer wavelength FIR, this extra contribution may be calibrated, at 
least for some classes of galaxies. Many project on the upcoming Herschel telescope will be 
devoted to the investigation of the evolved stars contribution to the FIR emission of galaxies. One of 
the extant questions is whether  the peak of the FIR emission (located in the 
wavelength range 70--100~$\mu$m) can be used as a reliable tracer of current SFR, and 
what limitations to its applicability may come from contamination of evolved populations. 

\acknowledgments
This work has been supported by the JPL, Caltech, Contract Number 1224667. It is part of 
SINGS, The Spitzer Infrared Nearby Galaxies Survey, one of the Spitzer Space
Telescope Legacy Science Programs..

\end{document}